\documentclass[article]{revtex4}
\usepackage{amsfonts,amsmath,amsthm}

\DeclareRobustCommand\openone{\leavevmode\hbox{\small1\normalsize\kern-.33em1}}%

\begin{document}
\title{\bf Cryptographic encryption scheme for solving the trusted courier problem based on metastable excited nuclei.}

\author{Thomas Durt}
\email{thomdurt@vub.ac.be}
\author{Alex Hermanne}
\email{aherman@vub.ac.be}

\affiliation{TONA Vrije Universiteit Brussel,
Pleinlaan 2, 1050 Brussels, Belgium}

\date{\today}
\maketitle


\noindent {\bf Abstract}: Quantum cryptography makes it possible to expand a short shared key (of e.g. 256 bits\cite{lange}) into an arbitrary long shared key. The novelty of quantum cryptography is that whenever
a spy tries to eavesdrop the communication he causes disturbances in the transmission of the message. Ultimately this unavoidable disturbance is a consequence of Heisenberg's uncertainty
principle that limits the joint knowledge of complementary observables. 

Now, a problem remains: in order to initialize quantum key distribution, Alice and Bob need to share a short shared key in order to be able to identify each other unambiguously. Therefore a trusted courier is needed. We propose in this paper a solution to the trusted courrier problem that was inspired by quantum cryptography. The idea is to encode the key that Alice sends to Bob into unstable nuclei in such a way that the message gets revealed only after the courrier has delivered it to Bob. 

In this approach, we replace Heisenberg
uncertainties by another type of uncertainty, that characterizes the knowledge of the time at which an unstable nucleus decays. As we shall show, this scheme makes it possible to refresh a key even in the case that we do not trust the courier who carries the key.


\section{Introduction} One of the main differences between quantum physics and classical physics is that the quantum theory is an intrinsically undeterministic theory. This has as a consequence that the knowledge that one can get about quantum systems is limited, as is expressed by Heisenberg uncertaintites that bound the simultaneous knowledge of conjugate quantitites such as position and momentum \cite{heis,rob}. These uncertainties also limit our ability to control and predetermine certain properties of quantum systems. For instance, it is impossible to prepare a photonic polarisation state in which the polarisations are simultaneously predetermined in the linear horizontal-vertical polarisation basis and in the circular polarisation basis, a feature that plays a key role in the BB84 protocol \cite{BB84}.

Similar uncertainties also limit time and energy \cite{wigner,booktime} with as a consequence that the exact time at which an unstable radioactive isotope will decay is characterized by an intrinsic uncertainty, of the order of the lifetime of the unstable isotope. 

Such isotopes present numerous medical applications such as PET scan tomography, cancer treatment and so on. As many other biomedical applications, they generated a specific industry with as a consequence that the technology in this field is very performant and efficient. One can now, at relatively low costs, select nearly {\it \`a la carte} the lifetime, decay mode and the chemical affinities of the radioactive substance that is produced in accelerators, depending on the requested application.

As the radioactive decay process is an intrinsically stochastic in time process that cannot be controlled from outside, one can also conceive, as we intend to show in the present paper, a key distribution protocol that exploits this limitation to the advantage of the authorized transmitters and receivers of the key (Alice and Bob). The basic idea is that if the lifetime of the radioactive isotope that is used to encode the fresh key is significatively longer than the transit time between them, the spy (Eve) cannot decipher the key because the decay will most often occur only after the courier has reached the receiver.

Of course the confidentiality of such a technique is based on the impossibility for Eve to eavesdrop the authorized users of the cryptographic transmission line outside of the travel period of the courier, but this basic assumption must be valid in the framework of classical and quantum cryptography as well.

The problem with classical encryption schemes is that nothing prevents Eve to learn the content of the key. She can always, in principle, mask her interaction afterwards by rewriting a fidel copy of the message and resending it to the receiver (Bob)\footnote{This is what is commonly called an intercept resend strategy.}.

In our case, the situation is different because as we shall see in the chapter devoted to security aspects,  a part of the message is revealed during the transit period but either it is ultimately discarded or it is too small to be useful to Eve, and the rest of the message, the raw key is revealed later, after it has reached Bob and is out of reach for Eve. 

Another feature of our protocol is that the raw key is a subset of the originally sent key. This does not menace the confidentiality of the transmission scheme provided the original key is devoided of meaningful information and consists of a random series of bits arbitrarily chosen by the sender.
After the exchange of the raw key, Alice and Bob are free to transmit to each other any kind of useful information making use of the Vernam pad \cite{vernam}, as in more conventional quantum cryptographic distribution schemes\footnote{As is well known, confidentiality is then absolute provided the length of the key equals the length of the message to transmit.}.

One could object that our approach is unefficient in order to send long keys, but it could be coupled to conventional Quantum Key Distribution schemes, in order to solve the trusted courier problem that unavoidably appears at the initialisation process of a secured quantum transmission when Alice and Bob check that they are well who they are supposed to be, and not a malignous Eve.
\section{Protocol for Key Distribution through Radioactive Encryption Process.}

\noindent Let us know describe in detail a protocol for key distribution in which the signal, a random bit, is encrypted with the help of radioactive atomic nuclei. This protocol (that we shall from now on call the Radioactive Encryption (R-E) protocol) consists of five distinct steps:

i) Alice produces a substance that contains unstable radioactive nuclei with a well-defined lifetime $\tau_D$. She dilutes it {\it ''homeopatically''} in order to arrive to a dilution level so low that the probability that (at least) one unstable nucleus is present in a standard volume (or sample) of say 1 mm$^3$ is of the order of $\mu$ with $\mu$ significatively smaller than one (of the order of 10 percent is a reasonable choice as we shall discuss in the next section devoted to security aspects of the protocol).

Alice also produces a ''twin'' substance or placebo that is entirely comparable to the radioactive one (same chemical ligants, same production scheme excepted that this twin substance is not radioactive). The simplest way to produce this twin substance is to stock the radioactive one during a time quite longer than the lifetime, so that its radioactivity has virtually vanished away.

It is important to note at this level that when an atom is excited in a collision with a highly energetic particle produced in a cyclotron two possible (des)excitation schemes can occur: either the atomic number and/or the mass number change during the process (which is then a transmutation process) or the nucleus keeps the same amount of neutrons and protons but gets excited to a metastable state that will decay afterwards within a time of the order of the life time. 
It is the second reaction scheme that interests us because when two atoms share the same atomic and mass numbers all electronic levels are the same and therefore the excited state is in the practice undistinguishible from the ground state by spectroscopic or non-invasive methods\footnote{The transmutation channel is not interesting for cryptographic purposes because it is possible experimentally to differentiate different isotopes of a same element by measuring isotopic displacements of their spectral rays. This has been measured for instance for the $Pb$ element \cite{godefroid2}. In parallel, an accurate theoretical description of the phenomenon has been achieved for instance in the case of light elements for which mass effects are dominant \cite{godefroid1}.}. In the case of metastable excitation on the contrary, it is only by waiting that the excited state decays and by measuring the products of desintegration that one knows whether the atom has been excited in the past. Moreover, there is no way to accelerate the desexcitation process because it involves nucleic degrees of freedom that are so complex that they are not controllable in the practice.

In an irradiation process transmutation processes and metastable excitations occur simultaneously but are characterized by different lifetimes and decay characteristics. 
Depending on which material is excited (targets of natural isotopic composition or isotopically enriched targets) and in which conditions (type and energy of bombarding particle, thickness of target) one can tune the relative weights of the "metastable decay channel" and of the "transmutation channel".  In appendix we provide an explicit example (the case of metastable  $^{117m}Sn$ isotopes) in order to show how one can selectively produce a metastable excited nucleus with a life time of the order of two weeks.

ii) Alice now divides a substrate plate into $M$ pairs of neighbouring cells with $M$ sufficiently high (at least say 1000) so that we enter the domain of validity of the law of large numbers.  For instance a standard A4 piece of paper contains 2400 squared cells of 5 mm side\footnote{We are currently investigating experimental limitations on the detection accuracy inherent to our protocol. This study is still in a preliminary phase and it is out of the scope of the present paper to describe the experimental aspects of our encryption scheme.}.  She spots a standard volume of radioactive substance on one cell chosen at random in each pair with the convention that say the left cell is attributed the binary value 1 and the right one the value 0. By doing so she generates a random series of $M$ bits that is physically encrypted on the plate. She systematically spots a standard volume of the neutral ''twin'' substance (placebo) on the remaining cell in each pair. She finally impregnates the whole plate with a low dilution ''neutral'' (placebo)  solution in order to avoid that Eve is able to localize ''empty'' cells where no nucleus at all of the isotopic specy that we use for encrypting the message would be present\footnote{By doing so, Alice could observe and memorize which are the pairs of cells in which only one cell contains the isotope that we use for encryption, which would occur with probability of the order of $2\cdot\mu\cdot (1-\mu)$. Afterwards, by listening to Alice-Bob communication she could guess a fraction of the order of ${ \mu\cdot (1-\mu)\over \mu}$ of the bits effectively measured by Bob, more or less, when $\mu\approx 10 \%$, $90 \%$ of them! When all cells are impregnated with low dilution placebo solution, there is a large number of non-excited nuclei in each cell so that their counting, cell by cell, is deprived of information that could appear to be useful for Eve.}.

iii) She sends the plate to Bob by use of a courier Charles. 


Charles can be physically concretized by a company that sends express mail, or a governmental mail company, or any person that travels from Alice to Bob. In conventional cryptographic schemes, Charles is the weak link of the protocol. Charles could get bribed, or the mail could get temporarily stolen during the trip without that he notices its disappearance. Eve could also intercept the letter in the mailbox of Bob before Bob opens it. 


iv) After having received the plate, Bob hides it during a time $\tau_B$ in a safe place where supposedly Eve has no access, an hypothesis that must be met whichever cryptographic protocol Alice and Bob choose to adopt otherwise it is no longer worth trying to establish secrecy at all. During that time Bob measures (for instance with a ccd detector or more simply by enveloping the plate into a photographic film) the occurence of decays which provides him a series of bits. Supposedly this is done with a spatial precision that is accurate enough in order to allow Bob to differentiate 0 and 1 bits.

The decay time $\tau_D$ must be chosen in order that the following constraints are met:

$\tau_D \leq \tau_B$ (the ''revelation'' time during which Bob waits that the message gets revealed must be larger than or comparable to the decay time otherwise too few secret bits get revealed to Bob).

Moreover it is reasonable to assume that the time $\tau_P$  that elapses between the production of the radioelement in the cyclotron and the encryption of the message by Alice, as well as the transportation time $\tau_T$ during which Charles carries the message from Alice to Bob are smaller than the decay time:

$\tau_P < \tau_D$  and $\tau_T < \tau _D$ (the expedition time is smaller than the decay time otherwise too many bits are lost underway to Bob).

Similar constraints are actually already fulfilled in the case of medical applications, where moreover $ \tau _D$ itself may not be too long in order to avoid to expose the person who absorbs the radioactive substance to a too high dosis of radiation.

v) Finally, Bob notes which bits appeared during the time $\tau_B$. This constitutes the raw key. Bob communicates to Alice on a public channel the list that contains the labels of all the pairs for which he observed  that a bit was revealed, without of course precising which bit it was (so to say without precising whether the location was ''right'' or ''left'').

At this level Alice and Bob share a key with a very high level of confidentiality. If certain errors appear (due for instance to an accidental collision of a piece of the photographic plate with a cosmic ray), it is still possible for Alice and Bob to suppress them thanks to classical techniques (reconciliation \cite{benetrob, benetal}). Also, one can estimate an upper bound on Eve's information as we shall see in the next section. Taking account of the existence of a non-zero but low probability that Eve possesses some information about the raw key Alice and Bob can in principle still let diminish this probability and reach an arbitrary level of confidentiality thanks to another classical post-treatment of the data called privacy amplification \cite{benetrob, benetal, benetmau}.

It is not our goal to describe these classical post-treatments here. They require an authentified but not necessarily confidential classical line of transmission between Alice and Bob\footnote{One could object that this authentified line in turn requires a trusted courier or at least a preliminary shared key and so on, but this kind of arguments applies to any scheme and is, in last resort, empty: it could be that Bob was replaced by a malignous clone at the nursery, we shall never know...}. They are well-known and are daily implemented in the framework of quantum cryptography.

\section{Security of the R-E protocol.}
Some stochasticity is present in the R-E protocol from the beginning because it is not realistic to assume that Alice can create standardized samples with exactly one unstable nucleus in each sample. 
This is comparable to the current situation in  quantum cryptography where high rate single photon sources, when they exist, are not of common use. Instead of sending exactly one photon at a time, in most realistic applications, the source is a damped laser source that sends pulses with a Poisson distributed population. In order to avoid repetition (redundancy) of the encoded information that could potentially open the door to menacing eavesdropping strategies (like the so-called beamsplitting \cite{benetal} or translucent attacks \cite{butler} that we shall consider next), the average number of photons by pulse is chosen to be low (of the order of 10 percent). 

It would be extremely difficult in practice to control exactly how many excited nuclei are present in each sample but the R-E protocol remains operational provided the distribution of the number of unstable nuclei by standard volume is stable throughout time. This requirement can be satisfactorily met in the practice because biomedical applications require a very accurate dosimetry of radioactive elements and obey high standards of quality and control, for obvious safety reasons. Without loss of generality we shall consider in the following that the distribution of the number of unstable nuclei is Poissonian so that $P(N)={ \mu^N\over N !}\cdot exp^{-\mu }$. In the case that the distribution is not exactly Poissonian it is still possible to adapt the protocol.

In any case, due to the fact that a mixing process occurs during the ''homeopatic'' dilution stage, we expect no gross departure from the Poisson distribution in the sense that  when the probability $P_0$ that a sample does not contain any radioactive nucleus is of the order of 90 percent, the probability that it contains one, two, three ... such nuclei is of the order of 10, 1, 0.1 ... percent. 

It is important that the dosimetric protocol is always realized in repeatable conditions so that one is able to calibrate properly the populations of empty samples, single nucleus samples, two nuclei samples ... once for all.

This guarantees that the redundancy of the encrypted information is limited: the probability that a bit contains more than one excited nucleus relatively to the probability that it contains at least one excited nucleus is of the order of ${1\over 2}\mu$, and the probability that a bit contains more than two excited nucleus relatively to the probability that it contains at least one excited nucleus is of the order of ${1\over 6}\mu^2$, close to 0.001 when $\mu\approx 0.1$.

It is then reasonable to assume that non-empty pulses only consist of singlet and pairs. In analogy with so-called beamsplitting \cite{benetal} or translucent attacks \cite{butler} where a spy taps an optically encrypted communication by beamsplitting the signal with an asymmetric beamsplitter, keeping only a very small fraction and resending the rest to Bob, Eve could measure a bit value without being noticed. This occurs whenever the information is encrypted in a pair of excited nuclei and that exactly one among the two excited nuclei decays during the transportation time $\tau_T$. Eve's strategy consists of keeping a memory of the single decays that occur during the transportation time and waiting until Bob publically communicate to Alice which bits he observed.  

The probability $P_{translucent}$ that this strategy is payful (which means that Eve's and Bob's bits coincide), with $\mu$ taken to be a small parameter, obeys the relation \begin{equation}\label{translucent} P_{translucent}\approx {\mu\over 2 !}\cdot 2\cdot (1-
exp(-{\tau_P+\tau_T\over \tau_D}))\cdot exp(-{\tau_P+\tau_T\over \tau_D}) \cdot (1-exp(-{\tau_B\over \tau_D})),\end{equation} (where we also assumed for simplicity that the probability that no decay occurs over the full duration of the protocol is very low. This probability is of the order of $exp(-{\tau_P+\tau_T+\tau_B\over \tau_D})$ and is small provided $\tau_P+\tau_T+\tau_B >\tau_D$.).

Let us now consider other possible attacks, making the most conservative hypotheses about Eve's abilities. The optimal strategy for her is to wait until the end of the travel and to eavesdrop the message just before Bob receives it. We shall assume that Eve is able to check whether a nucleus decayed by non-invasive methods,  with maximal efficiency ($\epsilon_{Eve}=1$) and in a very short time, and also to erase the substance on the plate and replace it by a radioactive sample without leaving trace of her intervention. We shall even assume that she possesses a ''bright'' source of excited nuclei meaning hereby that she can produce a standard volume of the substance with {\it at least} one excited nucleus in each sample in such a way that the relative population of single excited nuclei, pairs, triplets and so on exactly mimicks the relative populations produced by Alice (excepted  that, in the case of Alice, non-radioactive, neutral, samples are also produced with a probability close to 90 percent). 

This allows Eve to perform an analog of so-called opaque attacks where she replaces ''fainted'' bits by bright bits \cite{ben,butler} in order to increase the populations of bits that she intercepts-resends relatively to the population of bits that Bob receives. Of course by doing so Eve alters the statistics of the results and in order to mask her intervention different strategies are possible depending on the mechanisms of control exerted by Bob. There are two possible scenarios depending on the ability of Bob to check,  at the arrival of the message, whether a nucleus decayed underway.

a) If he can do so, Bob checks which nuclei are decayed at the arrival of the plate, before he waits that the other nuclei decay. For instance this can be done by enveloping the plate into a photographic film before Alice gives it to Charles. Bob can then develop the film and communicate to Alice the result of his observation by a public channel so that they discard those bits in the following.

The best strategy for Eve is then to replace some already decayed samples by ''bright'' samples that will decay with probability unity instead of $1-P_0\approx1-exp(-\mu)\approx 10 \%$. We shall even conservatively assumes that Eve can do so without affecting the relative populations of singlets, pairs, triplets and so on \cite{Durt}. The problem remains anyhow that by doing so Eve diminishes the number of already decayed samples that Bob measures when he receives the plate from Charles, and this could be noticed by Bob unless this departure agrees with the typical size of the statistical fluctuations that unavoidably occur in such circumstances. When the sample is long enough, these fluctuations obey the law of large numbers and their relative size decreases like the inverse of the square root of the lenght of the sample. Therefore, one can estimate an upper bound on the information possibly possessed by Eve. It is worth noting that even if Bob's non-invasive check is not perfectly accurate (for instance because the photographic plate did not react although a nucleus decayed, what could arrive with nonzero probability), one can still estimate, knowing what is the probability $\epsilon_{Bob}$ for Bob to reveal the trace of a decay process, a safe upper bound for Eve's information, making use of the law of large numbers (here we shall also conservatively assume that Eve's detectors are ideal: $\epsilon_{Eve}=1$). 

The reasoning goes as follows. The probability $P$ that a metastable nucleus decays before it arrives to Bob (where we also assumed for simplicity that the duration of the full protocol is large in comparison to $\tau_D$) is equal to $1-exp(-{\tau_P+\tau_T\over \tau_D})$. When the population of excited nuclei per encrypted bit is Poissonian of average $\mu$, that Bob's efficiency in revealing whether or not a nucleus has decayed is equal to $\epsilon_{bob}$ and that Alice sends one nucleus, the probability that it is observed by Bob is Poissonian with an average equal to $\epsilon_{Bob}\mu$. 

Let us assume that Alice sends $N$ bits which means that more or less $N\cdot \mu$ excited nuclei are sent, with $\mu$ taken to be a small parameter. The number of decayed nuclei that are present when Bob receives the message from Charles obeys a Bernouilli binomial distribution of mean $N\cdot \mu\cdot P$. Its variance is then of the order of $N\cdot \mu\cdot P\cdot (1-P)$. A safety threshold of five standard deviations ensures that Eve knows at most $5.\sqrt{N\cdot \mu \cdot P\cdot (1-P)}$ bits while Bob knows in average $N\cdot (1-P)\cdot (1-exp(-\epsilon_{Bob}\mu))\approx N\cdot (1-P)\cdot \epsilon_{Bob}\mu$ of them (those that decay after Bob receives the message). The probability $P_{intercept-resend}$ that Eve knows Bob's bit by such an intercept-resend strategy is thus (with a safety margin of five standard deviations) upperly bounded:

 \begin{equation}\label{intercept-resend}P_{intercept-resend}< {5\cdot \sqrt{ \mu P\over (1-P)\cdot(1-exp(-\epsilon_{Bob}\mu))^2}\over \sqrt{N}}\approx {5\cdot \sqrt{  P\over (1-P)\cdot \epsilon_{Bob}^2\mu}\over \sqrt{N}}.\end{equation}

b) It can be that Bob  does even not check whether a nucleus decayed underway by non-invasive methods (for one or another reason, for instance because it would be too expensive to do so) and has to wait during a time $\tau_B$ before he looks at the plate. Then Bob knows that, even in that case, Eve possibly knows at most a fraction $P$ of the bits, with  \begin{equation}\label{P}P\approx 1-exp(-{\tau_P+\tau_T\over \tau_D}), \end{equation} that she eventually replaced by bright bits\footnote{Actually in this case, Eve ought also to replace an equivalent fraction of Bob's bits by neutral bits in order to mask her intervention. This means, strictly speaking,  that instead of knowing a fraction $P$ of Bob's bits, Eve knows a fraction ${P\over 1-P}$ of them. Now, ${P\over 1-P}\approx P+P^2$ when $P$ is a small parameter, and this is a second order effect that we shall consistently neglect in the following.}.

Actually the opaque attacks and translucent attacks described above (\ref{intercept-resend},\ref{P},\ref{translucent}) are not fully independent but we can conservatively bound the probability that Eve knows a bit by the sum of the probability to know a bit by a translucent attack and the probability to know it by an opaque attack. Now, one can easily convince oneself that $P_{translucent}, P_{intercept-resend}$ and $P$, the three  upper bounds on the probabilities of succesful eavesdropping strategies by Eve that were estimated above, go to zero when ${\tau_P+\tau_T\over \tau_D}$ goes to zero, with $\tau_B$ comparable to or larger than $\tau_D$. 

In principle, Eve's information can thus be made arbitrarily small by letting increase $\tau_D$ and $\tau_B$, keeping $\tau_P$ and $\tau_T$ fixed or by letting decrease $\tau_P$ and $\tau_T$, freezing the values of $\tau_D$ and $\tau_B$. Of course those delays cannot be varied entirely at will, but there exists a large spectrum of metastable radio elements of which the time life varies from some hours to some days that could meet the constraints, depending essentially on the production and transportation times $\tau_P$ and $\tau_T$.

In any case the ratio between Eve's information and Bob's information about Alice's originally encrypted key (or rather about the subset of it that has not been discarded at the end of the process) is upperly bounded. Depending on the ability of Bob $\epsilon_{Bob}$ to check whether a nucleus decayed by non-invasive methods and of the choice of $\tau_D$, $\tau_B$, $\tau_P$ and $\tau_T$, this ratio can become small enough so that the reconciliation protocol and the process of amplification of privacy are succesful and finally allow Alice and Bob to generate a secret key with an arbitrarily high guarantee of confidentiality.
\section{Discussion and conclusions.}
One could object that our protocol is not specifically quantum. It is indeed so that nothing guarantees that the bits are encrypted in non-commuting bases (roughly speaking they are classical bits corresponding to the dichotomy excited (unstable)-neutral (stable)).

Now, in principle, classical bits can be perfectly cloned \cite{Dieks,Wootters} so that it is possible to intercept-resend them with a probability arbitrarily close to unity. If Eve had the opportunity to measure unambiguously whether a nucleus is excited or not she could intercept resend the signal with a probability close to unity without being noticed. If that was possible, this would mean that one could accelerate the decay process of unstable nuclei significatively in a relatively short time (of the order of one day). This would also mean that the problem of radio-active waste would be potentially solved. No progress was made in this direction during the last fifty years, although the environmental implications of such a progress would be overwhelming...As far as we know, to the contrary of excited electronic states, no stimulated radioactive decay is possible.

To formulate it clearly, our basic assumption is that one cannot accelerate the decay rate of radio-elements by any mean. At first sight this impossibility is seemingly more of technological nature than of fundamental nature in which case it could be overcome in the futute but as we shall argue now, nothing is less sure.

Firstly, it has been empirically confirmed that unstable quantum systems obey time-energy uncertainty relations (for instance the uncertainty in energy of a spectroscopic ray times its lifetime is always larger than Planck's constant).

Considered so the impossibility to force an unstable system to decay at a certain time would be due to its intrinsic quantum spread in time. We use the conditional form here because the status of time-energy uncertainty relations remains today a somewhat polemical and controversial subject \cite{booktime,arrival}.

Secondly, the security of our protocol can also be seen as a consequence of the fundamental impossibility to reach perfect control and/or predictability of a complex system.

Indeed, the temporal evolution of a metastable nucleus is a many body problem for which our knowledge is per se limited: we are unable to know exactly what is the initial state of the system; we have only partial knowledge about the interactions that occur when a proton collides a nuclei and also between the protons and neutrons inside the nuclei (it is commonly accepted that hadrons are bound states of triplets of quarks, but relatively few is known about the effective interaction that takes place between such three-body objects). To conclude, in order to be able to influence the behavior of such excited nuclei, we should be able to prepare them and to predict their evolution in time with nearly arbitrary accuracy, and also to interact with them in a nearly perfectly controllable way. 

Considered so our proposal presents a level of confidentiality that is intermediate between the one offered in conventional classical encryption schemes based on the length of the computational time necessary for breaking the key and the one offered in quantum cryptography that is based on the lack of predictability-measurability imposed by Heisenberg's uncertainties\footnote{We developed in the past \cite{probing} a protocol for key encryption for which the security is guaranteed by the imperfection of available sources and detectors. In that case too, we assumed that the ability of Eve to control/produce and/or detect single photons was {\it per se} limited.}.

Besides, our protocol shares some common points with the protocol proposed by Goldenberg and Vaidman 
in their paper of 1995 entitled Quantum Cryptography Based on Orthogonal States \cite{vaidman} in which although the bits are not encrypted in non-orthogonal states (non-commuting bases) the protocol is safe due to the fact that Eve has no simultaneous access to both components of the signal. During the time that the first component flies to Bob, the second component is stuck in Alice's lab; during the time that the second component flies to Bob, the first one is already safe in Bob's lab.

The security of Goldenberg-Vaidman's protocol was, in last resort, not a consequence of quantum uncertainties, but rather of causality. In other words, in order to break the code, Eve would have had to dispose of a time machine.

Notwithstanding the limitations imposed by time-energy uncertainties, in order to break our code, Eve would have to dispose of Maxwell's demons which like time machines seemingly belong to the domain of fantasy rather than to the real world...\footnote{According to A. Peres \cite{peres} for instance, the irreversibility of the measurement process, which is an essential ingredient of the no-cloning theorem \cite{Dieks,Wootters} is intrinsically linked to irreversibility in time in {\bf classical} mechanics.}

\subsection*{BB84 protocol with metastable states.}
If we consider metastable electronic states instead of radioelements, it is no longer so that Eve is unable to measure whether a state is excited or not in a time lapse shorter than the lifetime of the resonance, because in principle she could coherently guide the metastable state to an unstable state of quite shorter lifetime, with a laser field (induced transition) or with an electric field (quenching). Then she can measure in principle the photon that is emitted when this unstable state decays.

The essential difference between electronic states and nucleic states is the complexity of the system that explains why it is not possible, today, to accelerate the decay process of a nucleic metastable state.

Now, it could be that in the future one becomes able to manipulate nuclei as easily as electrons, in which case the safety of our protocol would be menaced. In such circumstances it is still possible to restore confidentiality by making use of the quantum degrees of freedom of the system. Let us denote $|\mathrm{\psi}^0\rangle$ the ground state of the nucleus and $|\mathrm{\psi}^1\rangle$ the excited metastable state.  We could encode in principle a qubit in a single nucleus, and use two bases of encryption, for instance the $\{|\mathrm{\psi}^0,\rangle$$|\mathrm{\psi}^1\rangle\}$ basis and the $\{|\mathrm{\tilde \psi}^0,\rangle$$|\mathrm{\tilde \psi}^1\rangle\}$ with $\{|\mathrm{\tilde \psi}^0={1\over \sqrt 2}(|\mathrm{\psi}^0\rangle+|\mathrm{\psi}^1\rangle)$ and $\{|\mathrm{\tilde \psi}^1={1\over \sqrt 2}(|\mathrm{\psi}^0\rangle-|\mathrm{\psi}^1\rangle)$. Formally such a protocol is equivalent to BB84 protocol \cite{BB84}, and in this case confidentiality is guaranteed by Heisenberg uncertainties.

The advantage of encrypting the signal in metastable states is the longevity of the encryption, which also means that nuclei can be seen as a quantum memory\footnote{Funnily, when we realize the BB84 protocol with long lived quantum states we are getting close to Wiesner's original idea, that can be traced back to the prehistory of quantum cryptography \cite{prehistory}, according to which one could create uncounterfeitable banknotes by numbering them thanks to quantum bits. It is not taken for granted however that bankers and postmen would accept to manipulate radioactive paper...}. It could provide an alternative way to refresh a key by quantum key distribution in the case that no conventional transmission line (optical fiber, aerial transmission or else) would be available.

It also brings an original solution to the ''trusted courier problem''.

\medskip                      
   \leftline{\large \bf Acknowledgment}
\medskip T.D. acknowledges support from the
ICT Impulse Program of the Brussels Capital Region (Project Cryptasc), the IUAP programme of the Belgian government, the grant V-18, and the Solvay Institutes for
 Physics and Chemistry.
\section*{Appendix: production of a metastable Sn isotope of halflife 13.6 days.}
By bombarding natural Sn (composed of nine stable isotopes with mass between 112 and 124) with high energy protons, many radioactive nuclides of the elements Sb, Sn, In and Cd with half lives longer than 1 min can be formed (for instance more than 50 when the energy of the protons is 65 Mev \cite{alex}). Excitation functions, expressing the probability of the production in function of the particle energy, for 13 among them, all emitting $\gamma$-rays with energy above 100 keV, are documented in Ref.\cite{alex}. One of these radionuclides is the metastable state $^{117m}Sn$ of the stable, naturally occurring, isotope 117Sn. The decay is characterized by a halflife $T _{1/2} = 13.6$ days\footnote{The halflife $T _{1/2}$ equals $| ln 2|$ times the mean life time $\tau_D$.} and the desexcitation from the metastable state to the ground state occurs through a cascade emission of a 156.0 keV $\gamma$-ray to an intermediate excited level followed by a 158.56keV $\gamma$-ray. 

These transitions are characterized by high internal conversion rates which mean that the energy of the $\gamma$-rays is converted to electronic excitation states with high probability, which in turn induces several X-lines with energies between 25 and 29 keV. 

The shape of the excitation function suggests that only a small fraction of the metastable state is formed by inelastic (p,pÕ) scattering on 117Sn while the largest part is due to (p,pxn) reactions on 118,119,120 Sn. The contribution of decay of 117Sb (only 0.07 $\%$ decays to 117$^mSn$) can be neglected. For incident protons of 30MeV (a standard value for commercially available isotope production machines) a thick target yield of 1MBq/$\mu$Ah can be expected\footnote{A thick target is such that all incomings protons are absorbed. It can be obtained with a metal plate of thickness of 1 millimeter. One Megabecquerel per micro Ampere hour means that when 
${10^{-6}\cdot 3600\over 1,6\cdot 10^{-19}}$ protons hit the target, a rate of $10^6$ desintegrations per second is generated. This corresponds to a number of metastable nuclei of the order of 
$10^6\cdot 13,6\cdot 24\cdot 3600 \cdot {1\over  ln 2 }$. So in a time of the order of an half-hour, of the order of $10^{12}$ radioactive bits could be in principle produced.}. 

Contaminating radio- or stable nuclides of Sb and In can be removed by a standard chemical separation step shortly after the end of bombardment. The production of the possibly disturbing co-produced Sn isotopes 113Sn ($T _{1/2}$ = 115.1 days); 119mSn ($T _{1/2}$ = 293 days), 121$^m$Sn ($T _{1/2}$ = 50 years) and 123Sn ($T _{1/2}$ = 129.2 days) can be limited or avoided by working with enriched 118Sn targets. Moreover, isotopes with a long lifetime would be present in the placebo substance in comparable proportions. Finally, it has to be remarked that for 117$^m$Sn production an alternative to charged particle activation is (n,$\gamma$) caption in a fission reactor on enriched 116Sn targets.

\end{document}